\documentclass[journal]{IEEEtran}

\usepackage{amsmath,amsfonts}
\usepackage{algpseudocode}
\usepackage{algorithm}
\usepackage{array}
\usepackage[caption=false,font=normalsize,labelfont=sf,textfont=sf]{subfig}
\usepackage{textcomp}
\usepackage{stfloats}
\usepackage{url}
\usepackage{verbatim}
\usepackage{graphicx}
\usepackage{multirow}
\usepackage{xcolor}
\usepackage{cite}

\title{A Negotiation-Based Multi-Agent Reinforcement Learning Approach for Dynamic Scheduling of Reconfigurable Manufacturing Systems}

\author{
  Manonmani Sekar$^{1}$, Nasim Nezamoddini$^{1}$ \\
  \thanks{
     $^{1}$ Industrial and Systems Engineering Department, Oakland University, Rochester, MI, USA. Emails: manonmanisekar@oakland.edu, nezamoddini@oakland.edu.
  }
}

\date{}

\begin{document}

\maketitle

\begin{abstract}

Reconfigurable manufacturing systems (RMS) are critical for future market adjustment given their rapid adaptation to fluctuations in consumer demands, the introduction of new technological advances, and disruptions in linked supply chain sections. The adjustable hard settings of such systems require a flexible soft planning mechanism that enables real-time production planning and scheduling amid the existing complexity and variability in their configuration settings. This study explores the application of multi agent reinforcement learning  (MARL) for dynamic scheduling in soft planning of the RMS settings. In the proposed framework, deep Q-network (DQN)  agents trained in centralized training learn optimal job–machine assignments in real time while adapting to stochastic events such as machine breakdowns and reconfiguration delays. The model also incorporates a negotiation with an attention mechanism to enhance state representation and improve decision focus on critical system features. Key DQN enhancements—including prioritized experience replay, n-step returns, double DQN, Huber loss, and soft target updates—further stabilize and accelerate learning. Experiments conducted in a simulated RMS environment demonstrate that the proposed approach outperforms baseline heuristics in reducing makespan and tardiness while improving machine utilization.The reconfigurable manufacturing environment was extended to simulate realistic challenges, including machine failures and reconfiguration times. Experimental results show that while the enhanced DQN agent is effective in adapting to dynamic conditions, machine breakdowns increase variability in key performance metrics such as makespan, throughput, and total tardiness. The results confirm the advantages of applying the MARL mechanism for intelligent and adaptive scheduling in dynamic reconfigurable manufacturing environments.
\end{abstract}


\section{Introduction}
Modern manufacturing systems are becoming increasingly complex due to the simultaneous need to address product variety, dynamic environments, reconfigurability, and real-time constraints \cite{koren1999reconfigurable}. The demand for mass customization requires flexible scheduling strategies that can handle fluctuations in market demand, frequent product changes, and unexpected machine failures \cite{mehrabi2000reconfigurable}. Reconfigurable Manufacturing Systems (RMS) were introduced as a paradigm shift from rigid and fixed manufacturing systems to flexible systems designed with modularity and adaptability to enable rapid structural and functional reconfigurations. Beyond equipment design, reconfigurability extends to scheduling and control strategies, which must deal with dynamic job arrivals, machine breakdowns, and frequent system-level changes. However, developing real-time scheduling approaches that take advantage of the distributed and adaptive nature of RMS while ensuring global performance of the system remained a significant challenge \cite{putnik2001reconfigurability}.

The researchers tried to present various mathematical models to find the optimal production scheduling and reconfiguration setting in RMS systems \cite{pansare2023reconfigurable}. The proposed models aimed at process planning \cite{ashraf2018configuration}, part and operation scheduling \cite{azab2015modelling}, line balancing \cite{delorme2024line}, and layout planning \cite{maganha2019layout} in these systems. Planning configurations using two stage techniques \cite{moghaddam2020configuration} and task to configuration assignments\cite{yelles2021reconfigurable} are two common approaches addressed by these models. Some researchers also target layout design and process scheduling at the same time \cite{barros2021approach}. The models range from single product single objective planning \cite{sabioni2021integrated} to more complex multi-objective part-family processes optimization \cite{kazemisaboor2022simulation}. Multi-objective models tried to find the best plans while optimizing conflicting objectives, such as modularity, completion time, and cost simultaneously \cite{haddou2018modularity}. Traditional models focused on small scale forms of RMS production planning with limiting assumptions such as defined arrival rate and high levels of equipment reliability \cite{dou2010optimisation}. To handle the increased computation expenses of large scale models, they also explored column generation technique \cite{cui2025configuration}, and various heuristics \cite{bensmaine2014new} and metaheuristics including genetic algorithm \cite{fan2022improved} and particle swarm optimization \cite{dou2021multi}. The traditional scheduling approaches proposed, such as heuristics and mathematical programming, often rely on centralized optimization in mostly static settings and suffer from limited scalability, high computational cost, and slow response under uncertainty \cite{imsetif2025job}. Their centralized nature is also incompatible with the distributed semi-autonomous structure of RMS, where machines are designed to act as independent units \cite{bi2010reconfigurable}. These limitations motivate the need for decentralized and adaptive scheduling mechanisms that align with the original principles of reconfigurability.

Reinforcement Learning (RL) provides a promising paradigm for sequential decision-making in dynamic environments of manufacturing systems such as RMS systems \cite{zhang2021review}. In this technique, agents learn policies through trial-and-error interactions with their surroundings \cite{sutton2000policy}. The integration of RL with deep neural networks, known as Deep Reinforcement Learning (DRL), has significantly extended its applicability to high-dimensional problems and breakthrough techniques such as Deep Q-Networks (DQN) \cite{mnih2015human}, Double Q-learning \cite{vanhasselt2016deep}, and dueling networks \cite{wang2016dueling}, as well as policy-gradient methods, including proximal policy optimization (PPO) \cite{schulman2017proximal},  demonstrated strong performance across complex domains such as job shop scheduling with potential disruptions and failures\cite{dai2020deep}. To increase the capabilities of this technique, multi-agent reinforcement learning (MARL) introduced new visions for distributed planning of complex systems through coordinated exploration \cite{lowe2017multi}. This trend was followed by value-based approaches such as Q MIX \cite{rashid2018qmix} and VDN \cite{sunehag2018value} by decomposing the joint action-value function into individual components. MARL techniques had different variations, including the entralized, decentralized and centralized-training-decentralized-execution paradigms \cite{wang2020multi}. Integration of attention mechanisms with MADRL has enabled more effective coordination and communication among agents \cite{iqbal2019actor}. This mechanism helps agents selectively focus on relevant information from other agents \cite{jiang2018learning}. Other variations such as Soft Actor Critics (SAC) improved scalability of these techniques in continuous actions spaces \cite{pu2021decomposed}. Researchers showed that even relatively simple algorithms like PPO can achieve strong performance in cooperative multi-agent games when combined with proper parameter sharing and centralized value functions \cite{yu2022surprising}.  Recent techniques also addressed limitations of early approaches for tracking value functions by integrating techniques such as transformation network \cite{son2019qtran} and hierarchical control \cite{mahajan2019maven}, and extending centralized policy gradients to factored multi-agent settings \cite{peng2021facmac}. These advances  significantly improved the scalability and performance of MARL in complex environments.

To align with the distributed control structure of manufacturing systems, researchers investigated the potential of applying MARL techniques in production planning and scheduling applications \cite{tang2022multi}. In MARL, each machine or resource can be modeled as an autonomous agent that learns production policies while interacting with others in cooperative or competitive settings. In this domain, applications such as dynamic job shop scheduling \cite{liu2021multi}, cloud manufacturing \cite{wang2021multi}, and reconfigurable systems \cite{zhang2023multi}, demonstrated enhanced adaptability and resilience compared to traditional centralized methods. The application of MARL techniques to manufacturing systems showed that this platform is capable enough to handle uncertainties of stochastic job arrivals and machine breakdowns \cite{wang2021multi}. Graph neural networks (GNN) were also integrated with these frameworks to model complex relationships, which is common in most manufacturing systems\cite{zhang2023graph}. The previous results also showed efficiency of such integration for handling disruptions in manufacturing settings\cite{liu2024dynamic}. For reconfigurable systems specifically, \cite{chen2022heterogeneous} developed heterogeneous MARL approaches for reconfigurable assembly systems, handling diverse agent capabilities and objectives. In distributed manufacturing contexts, federated MARL enabled collaborative learning across multiple manufacturing facilities while preserving data privacy \cite{li2023federated}. In most of the proposed MARL applications, effective exploration remains a significant challenge due to the exponential growth of the joint action space. Previous research also showed that self-supervised multi-agent auto-curriculum in these platforms can lead to emergent tool use and complex coordinated behaviors through self-play \cite{baker2020emergent}. Effective coordination in distributed environments requires negotiation to resolve conflicts over shared resources. Negotiation has long been studied in multi-agent systems \cite{rosenschein1994rules, parsons1998agents, jennings2001automated}, providing strategies for conflict resolution and cooperative agreements. Integrating negotiation with MARL enables agents to align individual incentives with system-level goals, dynamically adapt to environmental changes, and resolve resource conflicts \cite{shen2019multi}. 

Despite its promise, most MARL-based scheduling methods neglect negotiation, while negotiation-based approaches often remain rule-based and lack adaptability. Existing scheduling methods for RMSs still face major limitations such as lack of adaptability to dynamic changes, poor scalability in large-scale systems, limited ability to balance multiple objectives, and weak support for real-time decision-making. Moreover, most existing approaches employ job-centric agent formulations rather than machine-centric agents, which are more natural for systems where machines autonomously manage reconfiguration decisions. Existing methods in the literature often overlook the specific economic trade-offs involved in reconfiguration decisions, including setup costs, opportunity costs, and long-term strategic benefits. There is also limited work on integrating multi-timescale learning to handle both immediate scheduling decisions and long-term reconfiguration strategies. Finally, existing MARL approaches for manufacturing systems typically assume homogeneous agents, whereas reconfigurable manufacturing systems often involve planning heterogeneous machines with different capabilities and reconfiguration characteristics. This paper addresses the existing literature gap by developing an MARL framework that combines learning-based adaptability with negotiation-based coordination. The proposed framework models scheduler agents as autonomous learners that use an auction-based negotiation mechanism to resolve their conflicts and achieve their best production plan. The main contributions of this paper are listed below:
\begin{itemize}
     \item Presenting a novel machine-centric MARL framework for real-time scheduling of tasks in reconfigurable manufacturing systems.
     \item Integration of state-of-the-art techniques (PER, N-step returns, double DQN, Huber loss, soft target updates, state normalization) with an attention mechanism for enhanced state representation and robust learning.
    \item Proposing negotiation-based decision making for scalable multi-agent coordination that simultaneously optimizes immediate scheduling decisions and long-term reconfiguration strategies.
    \item Considering agent heterogeneity that accommodates diverse machine capabilities and reconfiguration characteristics.
    \item Evaluation of system's robustness under random machine breakdowns and reconfiguration time and analyzing their impacts on makespan, tardiness, and completion rate.    
\end{itemize}

The remainder of the paper is organized as follows: Section 2 describes the formulation of the problem, its objectives, assumptions, and requirements. Section 3 provides the theoretical foundations of the proposed MARL architecture and its integration with the negotiation mechanism. Section 4 describes case study setting and training performance followed by comprehensive experiments to test efficiency of the proposed scheduling mechanism in Section 5. The robustness analysis is presented in Section 6. Lastly, conclusions and potential avenues for future work are mentioned.

\section{Problem Modeling }

\subsection{Problem Description}

The research addresses dynamic scheduling in a reconfigurable manufacturing system so that machines can change their configuration to provide different services and varied production capacities. The system receives uncertain orders and is subject to potential failures, and it needs to find the best configuration setting for machines and jobs scheduling plan that can reduce costs and complete production in shortest time. In the proposed framework, the system is modeled as multi-agent setting in which machines are considered as independent agents, which must allocate dynamically arriving jobs in real time.

Key modeling challenges include:
\begin{itemize}
  \item {Dynamic Job Arrivals:} Jobs appear randomly with differing properties (product family, processing time, due date, priority).
  \item {Machine Reconfigurability:} Machines can alter their configurations to handle various job families, incurring associated time and cost penalties.
  \item {Partial Observability:} Each agent accesses only its own local state (status, configuration, ready time), a limited set of job queue information (top-$k$ jobs), and partial negotiation history.
  \item {Decentralized Control:} Each machine schedules jobs independently via learned policies, without full system state knowledge.
  \item {Coordination and Negotiation:} Agents must coordinate via negotiation to avoid conflicts and optimize global objectives.
\end{itemize}
The aim is to develop decentralized agent policies that maximize overall system performance under multiple competing objectives.

\subsection{Optimization Objective}

The scheduling problem in RMS settings must balance multiple conflicting performance criteria such as production efficiency, timeliness, costs, and resource utilization. To capture these trade-offs, we define a weighted multi-objective function:

\begin{equation}
\min Z = \alpha \cdot Z_{\text{makespan}} + \beta \cdot Z_{\text{tardiness}} + \gamma \cdot Z_{\text{setup}} + \delta \cdot Z_{\text{utilization}},
\label{eq:obj}
\end{equation}

where $\alpha, \beta, \gamma, \delta \geq 0$ and $\alpha + \beta + \gamma + \delta = 1$ are weight parameters reflecting decision-maker priorities.

\subsubsection{Makespan Minimization}
The makespan criterion aims to minimize the completion time of the last job:
\begin{equation}
Z_{\text{makespan}} = \max_{i \in \{1,\ldots,N\}} C_i,
\end{equation}
where $C_i$ denotes the completion time of job $i$.

\subsubsection{Weighted Tardiness}
Timely job completion is promoted by penalizing jobs that exceed their due dates:
\begin{equation}
Z_{\text{tardiness}} = \sum_{i=1}^{N} \rho_i \cdot \max(0, C_i - d_i),
\end{equation}
where $\rho_i$ is the priority weight and $d_i$ is the due date of job $i$.

\subsubsection{Set up Cost}
Set up time include machine set up time ($SP_j$) and time related to machine reconfiguration overhead:
\begin{equation}
\Delta_{jt} =  SP_j+\sum_{k_1,k_2 \in K} SR^{k_1 \rightarrow k_2}_j \cdot z_{j,k_1,k_2,t}
\end{equation}
where $SR^{k_1 \rightarrow k_2}_j$ is the time of transitioning machine $j$ from configuration $k_1$ to $k_2$, and $z_{j,k_1,k_2,t} = 1$ indicates that such a transition occurs at time $t$. The summation of these times are penalized in the objective function:
\begin{equation}
Z_{\text{setup}} =  \sum_{t=1}^{T_{\max}} \sum_{j=1}^{M} \Delta_{jt}
\end{equation}

\subsubsection{Resource Utilization Penalty}
To prevent under utilization of resources, we introduce a utilization-based regularization term:
\begin{equation}
Z_{\text{utilization}} = \sum_{j=1}^{M} \left(1 - \frac{\sum_{t=1}^{T_{\max}} u_{j,t}}{T_{\max}} \right)^2,
\end{equation}
where $u_{j,t} = 1$ if machine $j$ is active at time $t$, and $0$ otherwise.

The weighted formulation (\ref{eq:obj}) enables dynamic prioritization of performance measures. For example, assigning a larger $\alpha$ emphasizes throughput efficiency, while higher $\beta$ prioritizes due-date adherence. Similarly, increasing $\gamma$ discourages frequent machine reconfigurations, and adjusting $\delta$ promotes better resource utilization. This formulation provides a flexible optimization framework that aligns with the principles of RMS by supporting adaptive trade-offs based on real-time production requirements.

\subsection{Constraints in Reconfigurable Manufacturing Environment}

The feasible action space in the RMS is governed by machine availability, configuration compatibility, and negotiation rules. These are formally expressed as follows:

\subsubsection*{1) Machine Availability}
\begin{equation}
a_{i,j,t} \leq 1 - B_{i,j,t}, 
\quad \forall i \in \{1,\dots,N\}, \; j \in \{1,\dots,M\}, \; t
\end{equation}
 where $B_{j,t}=1$ if any job is assigned to machine $j$ at time $t$. Jobs can be assigned to idle machines.

\subsubsection*{2) Configuration Compatibility}
\begin{equation}
\Delta t  = T_{t}^{\text{Prd.}} - \frac{\tau_i}{\text{Eff}(j)} - R_{i,j,t}, 
\end{equation}
with
\[
R_{i,j,t} = 
\begin{cases}
0, & \text{if } C_{j,t} = J_{i,t}^{\text{family}}, \\
\text{ReconfigTime}(i,j), & \text{otherwise}.
\end{cases}
\]

where $\tau_i$ is the process time of the job $i$, \text{Eff}(j) is the efficiency of the machine $j$ and $T_{t}^{\text{Prd.}}$ is the length of the production period $t$. the reconfiguration time is considered only if the machine is not in the correct configuration for the job family $i$. The assignment of the job is feasible only if $\Delta t\geq 0$.

\subsubsection*{3) Limited Job View}
A machine can only see and bid on a small number of
jobs defined in its view $V$ at a time.
\begin{equation}
B_{i,j,t} = 0, \quad \forall i \notin Q_t^{\text{view}}, \quad |Q_t^{\text{view}}| \leq V   .
\end{equation}

\subsubsection*{4) Single Assignment and Negotiation}
\begin{equation}
\sum_{j=1}^M B_{i,j,t} \leq 1, \quad \forall i,t,
\end{equation}
ensuring each job $i$ is assigned to at most one machine per step.  
If multiple machines compete for job $i$, a negotiation mechanism selects a winner $j^*$:
\[
B_{i,j^*,t}=1, \quad B_{i,j,t}=0 \; \forall j \neq j^*.
\]

\section{Proposed Methodology}



The dynamic scheduling problem in reconfigurable manufacturing systems can first be formalized under the framework of a Multi-Agent Markov Decision Process (MMDP):
\[
\text{MMDP} = (N, S, A, P, R, \gamma),
\]
where $N^{\text{agent}}=\{1,2,\ldots,n\}$ is the set of agents including jobs and machines, $S = S_1 \times S_2 \times \cdots \times S_n \times S_{\text{env}}$ is the joint state space including system-level information, and $A = A_1 \times A_2 \times \cdots \times A_n$ is the joint action space. The transition model $P(s'|s,a)$ captures the global dynamics, and $R$ defines a collective reward function to promote cooperation across all agents.

\subsection{Multi-Agent Reinforcement Learning Framework}

To solve the scheduling task under uncertainty and decentralized observations, we adopt a cooperative MARL formulation. The system is modeled as a cooperative Markov game, also referred to as a Decentralized Partially Observable Markov Decision Process (Dec-POMDP). The components are:

\begin{itemize}
    \item {Agents:} A set of jobs $i = \{1, 2, \dots, N\}$, A set of reconfigurable machines $j = \{1, 2, \dots, M\}$, A negotiation agent, A DQN scheduler agent.
    \item {Local Observations:} Each agent $n$ perceives only partial information $O_{n,t} \in \mathcal{O}_n$ such as its current workload and immediate neighborhood.
    \item {Action Spaces:} $\mathcal{A}_n$ for each agent, consisting of task assignment, bidding, or negotiation decisions.
    \item {Transition Model:} $P: \mathcal{S} \times \mathcal{A}_1 \times \cdots \times \mathcal{A}_{N^{\text{agent}}} \rightarrow \Delta(\mathcal{S})$, where $\Delta(\mathcal{S})$ denotes probability distributions over next states.
    \item {Reward Structure:} Individual rewards $R_i$ reflect job-level objectives, while the team reward encourages system-wide optimization.
\end{itemize}

At each time step $t$, the environment is in state $S_t \in \mathcal{S}$, while each agent $i$ observes $O_{n,t}$ and selects an action
\begin{equation}
A_{n,t} \sim \pi_n(A_{n,t} | O_{n,t}; \theta),
\end{equation}
where $\pi_n$ is the policy parameterized by $\theta$. In our framework, parameter sharing is employed such that $\pi_n = \pi$ for all $n \in \mathcal{N^{\text{agent}}}$, which improves coordination and sample efficiency.

The joint action
\begin{equation}
A_t = (A_{1,t}, A_{2,t}, \dots, A_{N^{\text{agent}},t}),
\end{equation}
determines the state transition
\begin{equation}
S_{t+1} \sim P(S_{t+1} | S_t, A_t).
\end{equation}
Each agent then receives an individual reward $R_{i,t}$, and a collective reward signal is defined as
\begin{equation}
R_t = \frac{1}{N^{\text{agent}}}\sum_{n=1}^{N^{\text{agent}}} R_{n,t}.
\end{equation}

The reward function guides the DQN agent towards optimal job scheduling by balancing multiple objectives listed in the Equation \ref{eq:obj}. 
The total reward is clipped between $R_{\min}$ and $R_{\max}$ to ensure training stability. The learning objective is to optimize the shared policy parameters $\theta$ to maximize the expected discounted return:
\begin{equation}
J(\theta) = \mathbb{E}_{\tau \sim \pi_\theta}\left[ \sum_{t=0}^\infty \gamma^t R_t \right],
\end{equation}
where $\tau$ denotes a trajectory generated by the policy $\pi_\theta$. This formulation ensures that the MARL framework simultaneously balances local agent incentives with the global scheduling objective, making it well-suited for dynamic, large-scale, and uncertain manufacturing systems.

\subsection{Centralized Training with Decentralized Execution }

The Centralized Training with Decentralized Execution (CTDE) paradigm addresses the core challenge in cooperative MARL that agents must learn coordinated strategies using global information during training while operating independently with only local observations. This is achieved by separating the training and execution phases.

\subsubsection{Centralized Training}
During training, a centralized critic is employed with access to the global state $S_t$ and joint action $A_t$. The critic estimates the expected return:
\begin{equation}
V^{\text{cent}}(S_t; \phi) \approx \mathbb{E}_{\pi_\theta}\left[ \sum_{k=0}^\infty \gamma^k R_{t+k+1} \,\Big|\, S_t \right],
\end{equation}
where $\phi$ are critic parameters. The policy network is updated using a centralized advantage:
\begin{equation}
\nabla_\theta J(\theta) = \mathbb{E}\left[ \nabla_\theta \log \pi(A_{n,t}|O_{n,t};\theta) \cdot A^{\text{cent}}_t(\phi) \right],
\end{equation}
with
\begin{equation}
A^{\text{cent}}_t(\phi) = R_t + \gamma V^{\text{cent}}(S_{t+1};\phi) - V^{\text{cent}}(S_t;\phi).
\end{equation}
This setup improves variance reduction, credit assignment, and coordination during learning.

\subsubsection{Decentralized Execution}
At execution, each agent selects actions using  its local observation:
\begin{equation}
A_{n,t} \sim \pi(A_{n,t}|O_{n,t}; \theta^*),
\end{equation}
where $\theta^*$ are the trained policy parameters. The centralized critic is discarded, ensuring scalability, robustness under partial observability, and low computational overhead. Coordination emerges implicitly from training and through negotiation mechanisms during interaction.
\subsection{Deep Q-Networks and Enhanced Variants}

The Deep Q-Network (DQN) introduced by Mnih et al.~\cite{mnih2015human} approximates the optimal action-value function
\begin{equation}
Q^*(s,a) = \max_\pi \mathbb{E}\!\left[\sum_{k=0}^\infty \gamma^k r_{t+k} \;\middle|\; s_t=s, a_t=a, \pi \right]
\end{equation}
using a neural network $Q(s,a;\theta)$. Learning proceeds by minimizing the temporal-difference (TD) loss
\begin{equation}
\mathcal{L}(\theta) = \mathbb{E}_{(s,a,r,s')}\!\left[\big(r+\gamma \max_{a'}Q(s',a';\theta^-) - Q(s,a;\theta)\big)^2\right]
\end{equation}
where $\theta^-$ are target network parameters periodically updated to stabilize learning. DQN further employs experience replay to break correlations between sequential transitions. Despite its success in Atari and other benchmarks, the vanilla formulation is limited by overestimation bias due to the $\max$ operator, inefficient $\epsilon$-greedy exploration, and poor feature representation when facing high-dimensional or structured states. 
\begin{figure}[htbp]
    \centering   \includegraphics[width=1.02\linewidth]{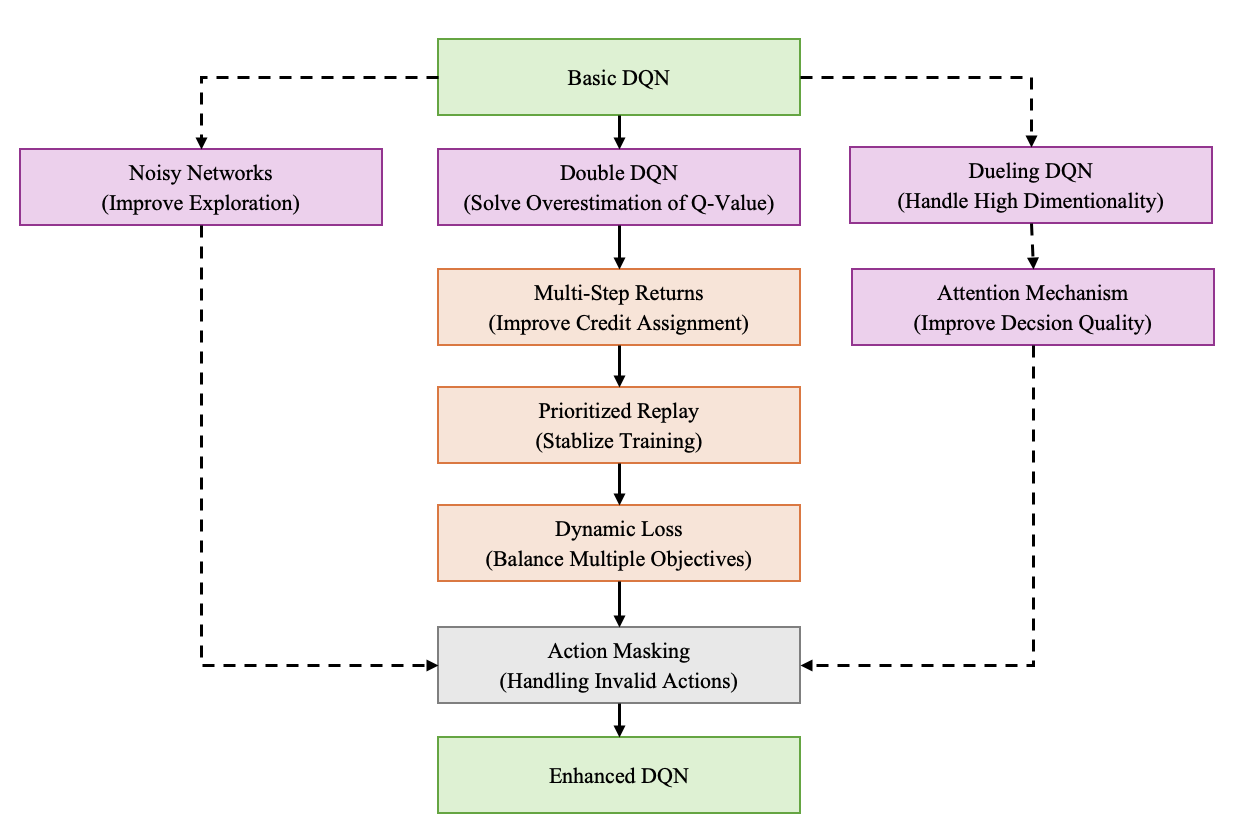}
    \caption{Enhanced DQN Architecture}
    \label{fig:fig1}
\end{figure}

We proposed Enhanced DQN presented in Fig \ref{fig:fig1} with following series of architectural improvements:
\subsubsection{Overestimation of Q-values} Standard DQN tends to overestimate expected rewards, which may lead to suboptimal policies. Double DQN ~\cite{van2016deep} mitigates this issue by using the target network to evaluate the action selected by the policy network, producing more accurate Q-value estimates:
\begin{equation}
Q_{\text{target}} = r + \gamma Q\!\left(s', \arg\max_{a'}Q(s',a';\theta);\theta^-\right)
\end{equation}

\subsubsection{Large State and Action Spaces} Real-world manufacturing involves numerous machines, jobs, and operation types, resulting in very large state and action spaces. Standard DQN architectures may struggle in such high-dimensional settings. Enhancements such as Dueling Networks~\cite{wang2016dueling} separate the value of a state from the advantage of taking specific actions, improving learning efficiency in large action spaces: 
\begin{equation}
Q(s,a) = V(s) + \left(A(s,a)-\tfrac{1}{|\mathcal{A}|}\sum_{a'} A(s,a')\right)
\end{equation}

\subsubsection{Exploration Strategy} Effective exploration is critical in complex environments. Noisy networks~\cite{fortunato2017noisy} replace $\epsilon$-greedy exploration with parametric noise in network weights, enabling more efficient and state-dependent exploration:
\begin{equation}
\theta = \mu + \sigma \odot \epsilon, \quad \epsilon \sim \mathcal{N}(0,1),
\end{equation}
Negotiation-based exploration can further allow context-aware exploration in multi-agent or constrained settings. 
\subsubsection {Correlated Experiences} Sequential experiences in manufacturing are often correlated, and random sampling from a standard replay buffer may destabilize training. Prioritized Experience Replay (PER)~\cite{schaul2016prioritized} improves efficiency by sampling transitions with higher temporal-difference (TD) errors more frequently, allowing the agent to focus on informative or surprising experiences:
\begin{equation}
P(i)=\frac{|\delta_i|^\alpha}{\sum_k|\delta_k|^\alpha}, \quad w_i=\Big(\tfrac{1}{N\cdot P(i)}\Big)^\beta
\end{equation}
\subsubsection {Temporal Credit Assignment} Identifying which action contributed to a delayed reward, such as job completion, is difficult in long horizons. n-step returns address this challenge by propagating cumulative rewards over multiple steps, providing a richer learning signal and improving temporal credit assignment:
\begin{equation}
G_t^{(n)} = \sum_{k=0}^{n-1}\gamma^k r_{t+k} + \gamma^n \max_a Q(s_{t+n},a;\theta^-)
\end{equation}
\subsubsection {Complex State Information} Manufacturing states may encode diverse features such as machine conditions, job queues, and job progress. Attention mechanisms can be employed to focus the agent on the most relevant state components, improving decision quality in environments with complex or large state representations:
\begin{equation}
\text{Attn}(Q,K,V)=\text{softmax}\!\left(\frac{QK^\top}{\sqrt{d_k}}\right)V
\end{equation}
\subsubsection {Handling Invalid Actions} In scheduling, not all actions are valid at every step, such as assigning jobs to busy machines. Action masking prevents the agent from attempting infeasible actions and ensures learning is directed towards valid scheduling decisions:
\begin{equation}
Q'(s,a)=
\begin{cases}
Q(s,a), & a \in \mathcal{A}_{\text{valid}}(s), \\
-\infty, & \text{otherwise}.
\end{cases}
\end{equation}
This is a mechanism designed to prevent actions for machine $j$ which are infeasible based on given constraints. Infeasible actions receive $-\infty$ not to be chosen again.
\subsubsection {Competing Objectives} The decision making in job scheduling problem includes competing objectives. Dynamic loss weighting balances these competing objectives by adjusting coefficients according to gradient norms~\cite{kendall2018multi}:
\begin{equation}
\mathcal{L}_{\text{total}} = \lambda_Q \mathcal{L}_Q + \lambda_U \mathcal{L}_U + \lambda_{\text{reg}}\|\theta\|^2.
\end{equation}
\subsubsection {Training Stability} Deep reinforcement learning is prone to instability. Several techniques improve robustness: Huber loss (smooth L1) reduces sensitivity to outliers compared to mean squared error, soft target updates (Polyak averaging) provide gradual updates for stable learning, and state normalization ensures features remain on a consistent scale.

By incorporating the above modifications, the enhanced DQN agent becomes more capable of handling the intricacies of realistic manufacturing scheduling problems, achieving more stable training and improved performance compared to basic DQN implementations or heuristic baselines. The overview of the enhanced DQN is presented in Algorithm \ref{alg:ctde_dqn}. Comparison between Baseline DQN and Enhanced DQN is shown in Table \ref{tab:qualitative-comparison}


\begin{table}[htbp]
\centering
\caption{ Comparison between Baseline DQN and Enhanced DQN}
\label{tab:qualitative-comparison}
\begin{tabular}{|p{0.15\linewidth}| p{0.3\linewidth}| p{0.3\linewidth}|}
\hline
Aspect & Baseline DQN & Enhanced DQN \\
\hline
Network architecture & Simple feed-forward multi-layer perceptron & Dueling DQN with attention and deeper feature extractor \\
\hline
Exploration strategy & Epsilon-greedy & NoisyNet + negotiation-guided proposals (hybrid) \\
\hline
Target update & Hard copy (periodic) & Soft (Polyak) updates (tau) \\
\hline
Replay buffer & Uniform replay & Prioritized Experience Replay (PER) with importance sampling \\
\hline
Return estimator & l-step TD & l-step returns (e.g. $l=3$) for better credit assignment \\
\hline
Loss function & MSE (L2) & Huber / Smooth-$L_1$ for robust TD updates \\
\hline
State preprocessing & Raw state & Running mean/std normalization (online) \\
\hline
Stability tricks & Basic & Gradient clipping, LR scheduler, PER warmup, larger batch sizes \\
\hline
Guided exploration & None & Simple negotiation agent proposes job-machine pairs (heuristic) \\
\hline
Intended benefits & Baseline learning & Faster credit propagation, better exploration, more stable updates \\
\hline
\end{tabular}
\end{table}

\subsection{Negotiation Mechanism }
\label{sec:negotiation}

\begin{figure*}[!htbp]
    \centering
    \includegraphics[width=0.8\textwidth]{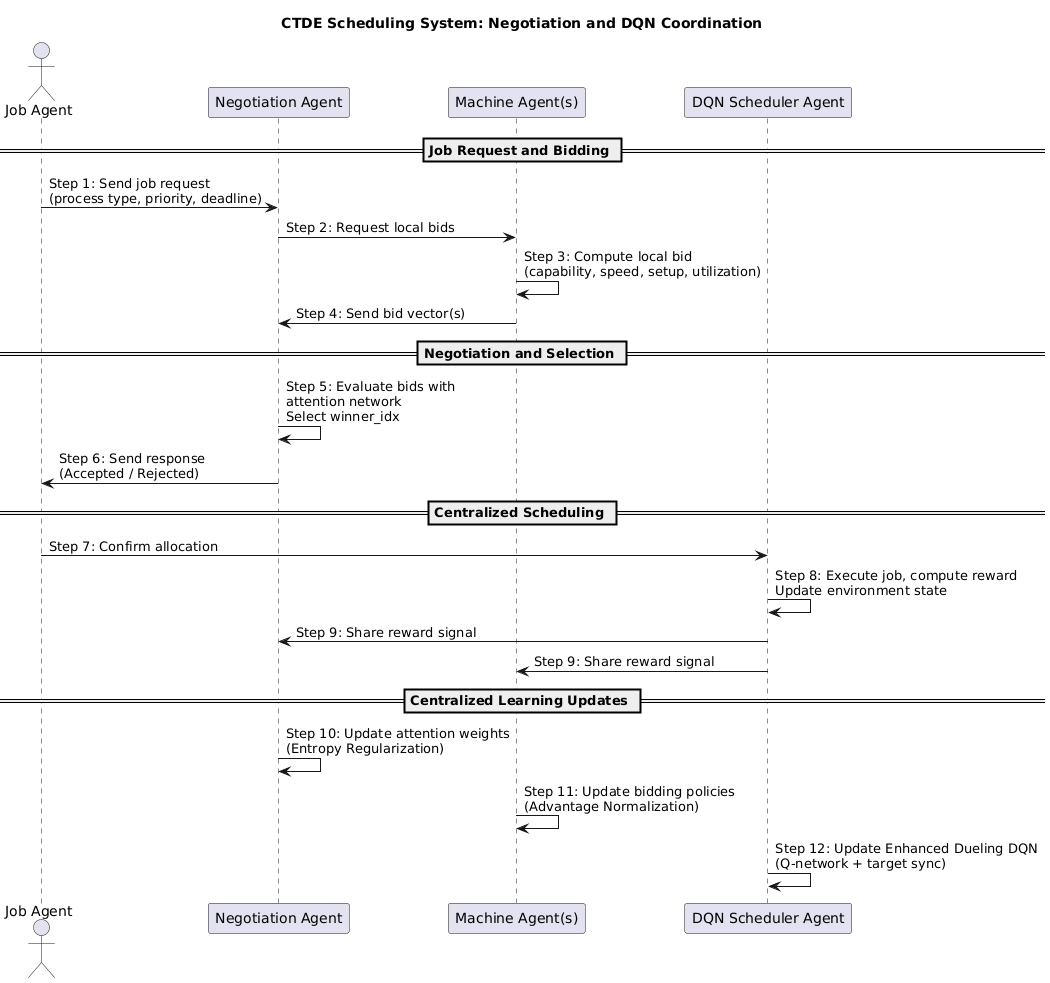}
    \caption{
        Negotiation process flow among Job, Negotiation, Machine,and DQN Scheduler Agents under the CTDE paradigm.The Negotiation Agent centrally evaluates machine bids using an attention network, while the DQN Scheduler performs global reward-based updates.
    }
    \label{fig:neg_flow}
\end{figure*}

The integration of a negotiation mechanism allows agents to explore cooperative decisions and make the enhanced DQN more stable, efficient, and better suited to complex resource-allocation problems such as reconfigurable manufacturing systems. 
In the proposed RMS scheduling framework, this negotiation mechanism serves as the interaction layer between job agents and machine agents with a DQN scheduler agent supervising the negotiations. The goal of the proposed scheduling framework is to dynamically assign jobs to machines in a distributed fashion based on current capabilities and process requirements. The integrated process is inspired from multi-side auctions when multiple buyers are matched with multiple sellers \cite{wurman2001parametrization}.During training, all participating agents share their experiences centrally while maintaining decentralized policies at runtime. The overall process is summarized in Figure~\ref{fig:neg_flow}. The bidding process is centrally managed by a
DQN scheduler agent and orchestrated by a negotiation agent to facilitate communication between job agents and
machine agents. The core workflow begins with the initialization of the global system state $ s_t = \langle J_t, M_t, Q_t, H_t \rangle $, encapsulating job queues, machine configurations, queue data, and historical context. Each negotiation starts when a job agent submits a job request vector $\mathbf{x}_i$ to the negotiation agent:
\begin{equation}
    \mathbf{x}_i = [p_i, d_i, \rho_i, \tau_i],
\end{equation}
where $p_i$ denotes the type of process, $d_i$ is the remaining time to the deadline, $\rho_i$ is the priority of the job, and $\tau_i$ represents the required processing time. Issuing a job request is followed by bid generation, evaluation, and confirmation by the DQN scheduler.
The negotiation agent requests local bids from all machine agents, where each machine agent $j$ computes  a local bid for a received job $i$
request:
\begin{equation}
    \mathbf{y}_{ij} = [f_j, r_j, u_j, c_{ij}, \Delta_{ij}],
\end{equation}
with 
$f_j$ its flexibility for adjustment, 
$r_j$ the reliability score,
$u_j$ the current utilization, $c_{ij}$ the estimated setup time, and
$\Delta_{ij}$ the expected processing cost for job $i$ on machine $j$. Each machine agent submits bids after checking its capabilities based on current production capability, as well as its potential capability with reconfiguration. 
In the proposed framework, the negotiation agent acts as a centralized mediator that receives bids from all machine agents and computes attention weights to determine the most appropriate machine for allocation. Its scoring network is parameterized as:
\begin{equation}
    h(\mathbf{y}_{ij}) = 
    \sigma \!\left( \mathbf{W}_2
    \, \text{ReLU}(\mathbf{W}_1 \mathbf{y}_{ij} + \mathbf{b}_1)
    + \mathbf{b}_2 \right),
\end{equation}
with learnable parameters $\{\mathbf{W}_1, \mathbf{W}_2, \mathbf{b}_1, \mathbf{b}_2\}$. The normalized bid scores are calculated using attention mechanism:
\begin{equation}
    \alpha_{ij} =
    \frac{\exp(h(\mathbf{y}_{ij}))}{\sum_{j=1}^{M} \exp(h(\mathbf{y}_{ij}))}.
\end{equation}
The machine with the highest attention weight $\alpha_{ij^*}$
is selected for assignment:
\begin{equation}
    m^* = \arg\max_{j} \, \alpha_{ij}.
\end{equation}

The negotiation agent then returns a decision to the job agent:
\[
\text{response} = 
\begin{cases}
    \text{Accepted}~(B_{i,j,t}=1), & \text{if $j=m^*$} \\
    \text{Rejected($B_{i,j,t}=0)$}, & \text{otherwise}
\end{cases}
\]
If the bid with reconfiguration is accepted, the machine will update its capability profile. Machine agents improve their bidding strategies and learn centrally through a shared loss function with advantage normalization and entropy regularization. Similarly, the negotiation agent learns the best scoring strategies using policy gradients on negotiation loss. The DQN scheduler agent represents the global decision maker and learns a global Q function evaluating job-machine assignments. It uses an {Enhanced Dueling Deep Q-Network} to estimate action-value functions for job–machine assignments:
\begin{equation}
    Q(s, a; \theta) = V(s; \theta) +
    \big(A(s, a; \theta) - \frac{1}{|\mathcal{A}|}\sum_{a'}A(s, a'; \theta)\big),
\end{equation}
where $V(s)$ and $A(s,a)$ denote the value and advantage streams, respectively.
The scheduler agent learns a global policy $\pi_D$ that maximizes the long-term cumulative reward:
\begin{equation}
    J(\theta) = \mathbb{E}\!\left[\sum_{t=0}^{T}\gamma^t r_t\right],
\end{equation}
with discount factor $\gamma$.
It also propagates global reward signals $r_t$ to the Negotiation and Machine Agents for coordinated training.
Once accepted, the DQN scheduler agent updates its policy using the global reward $r_i$ from the environment to improve long-term performance. Both negotiation and machine agents update their local parameters using the shared reward signal. To ensure stable coordination and exploration, each agent is trained with entropy-regularized policy gradients. During training, the gradients are computed centrally and distributed to each agent for decentralized execution. This hybrid structure allows
the system to scale efficiently to complex reconfigurable
manufacturing environments while maintaining coordinated
decision-making and adaptability.

When resource conflicts arise (e.g., multiple jobs target the same resource), a global attention mechanism is invoked to resolve them by computing priority-weighted urgency scores. Agents revise their bids until all conflicts are resolved and final allocations are confirmed. Following the assignment or the completion of the negotiation cycle, the Negotiation Agent executes a crucial {Learning} phase by updating its attention parameters through backpropagation. Negotiation experiences are recorded as tuples $ \langle s_t, a_t, r_t, s_{t+1} \rangle $ and used to update each agent's policy through DRL methods. This integrated mechanism ensures the system continuously improves its scheduling and resource allocation decisions based on past negotiation results, manifesting negotiation strategies learned with global information as decentralized bidding behaviors during execution.

\section{Case Settings and Training}
\label{sec:results}
The proposed reconfigurable manufacturing scheduling system is tested in a case study with the settings explained in this section. All experiments are conducted in the custom-developed {Enhanced Reconfigurable Manufacturing System (EnhancedRMS)} environment implemented in Python~3.8. 
The system was designed to emulate job–machine interactions, dynamic setup operations, and reconfiguration behaviors under stochastic conditions. Each simulation episode involved randomly generated job sequences with processing times, due dates, and setup requirements derived from uniform and normal distributions based on realistic Festo Didactic workstation data. 

\subsection{System Configuration}
The experimental testbed consists of a flexible job shop manufacturing system with the following characteristics:

\begin{itemize}
    \item {Machines:} 5 machines with varying capabilities
    \item {Jobs:} 50 jobs with sequential process requirements
    \item {Process Types:} 6 different process categories
    \item {Breakdown Scenario:} Machine 3 failure at simulation start
\end{itemize}

\subsection{Machine Characteristics}
Each machine possesses two distinct capability sets:
\begin{enumerate}
    \item {Native Processes:} 2-3 processes executable without reconfiguration
    \item {Reconfigurable Processes:} 2-3 additional processes accessible after reconfiguration (setup time: 3-7 time units)
\end{enumerate}

Table~\ref{tab:machine_config} presents a sample machine configuration.

\begin{table}[htbp]
\caption{Sample Machine Configuration}
\begin{tabular}{|c|c|c|c|c|}
\hline
{Machine} & {Native} & {Reconfigurable} & {Setup} & {Negotiation} \\
{ID} & {Processes} & {Processes} & {Time} & {Flexibility} \\
\hline
M0 & \{0, 2\} & \{1, 3\} & 5.2 & 0.85 \\
M1 & \{1, 3, 4\} & \{0, 2\} & 4.8 & 0.92 \\
M2 & \{2, 5\} & \{0, 1, 4\} & 6.1 & 0.78 \\
M3 & \{0, 3, 5\} & \{2, 4\} & 3.9 & 0.88 \\
M4 & \{1, 4\} & \{0, 3, 5\} & 5.7 & 0.81 \\
\hline
\end{tabular}
\label{tab:machine_config}
\end{table}

\subsection{Job Characteristics}
Jobs are characterized by:
\begin{itemize}
    \item {Process Sequence:} 3-5 sequential processes per job
    \item {Processing Time:} 5-15 time units per process
    \item {Priority Level:} 1-5 (integer scale)
    \item {Due Date:} 2-4× total process time
\end{itemize}

\subsection{Hyper-Parameters Settings}

Hyper-parameters were selected through automated tuning and the best configuration is summarized in Table~\ref{tab:bestconfig}.

\begin{table}[h]
\centering
\caption{Best Hyperparameter Configuration for Dueling-Attn-DQN}
\label{tab:bestconfig}
\begin{tabular}{l c}
\hline
{Parameter} & {Value} \\
\hline
Learning rate ($lr$) & 0.0001 \\
Discount factor ($\gamma$) & 0.99 \\
Hidden dimension ($h$) & 128 \\
Batch size ($B$) & 64 \\
Epsilon decay ($\epsilon_{decay}$) & 0.995 \\
Soft target update ($\tau$) & 0.005 \\
\hline
\end{tabular}
\end{table}

Training lasted for 1000 episodes using a replay buffer of capacity $5\times10^{4}$.  
The loss function was mean squared error (MSE), optimized with Adam. All runs were executed on a GPU-enabled PyTorch environment.

\begin{algorithm}[!htbp]
\caption{CTDE-Based Reconfigurable Manufacturing Scheduling with Enhanced Dueling DQN}
\label{alg:ctde_dqn}
\begin{algorithmic}[1]
\Require Number of jobs $N_J$, number of machines $N_M$, process types $\mathcal{P}$, 
episodes $E$, replay buffer $\mathcal{D}$, learning rates $\alpha_D$, $\alpha_N$, $\alpha_M$
\Ensure Trained Enhanced DQN agent $\pi_D$, Negotiation agent $\pi_N$, Machine agents $\{\pi_M^i\}$

\State Initialize Enhanced Dueling DQN $\pi_D$ with parameters $\theta_D$ and target network $\theta_D'$
\State Initialize Negotiation agent $\pi_N$ with parameters $\theta_N$
\State Initialize each Machine agent $\pi_M^i$ with parameters $\theta_M^i$
\State Initialize replay buffer $\mathcal{D}$ and negotiation buffer $\mathcal{B}$
\State Initialize environment $\mathcal{E}$ with job set $\mathcal{J}$ and machine set $\mathcal{M}$ having random capabilities

\For{each episode $e = 1, 2, \dots, E$}
    \State Reset environment $\mathcal{E}$, set $t \gets 0$, total reward $R_e \gets 0$
    \For{each job $j \in \mathcal{J}$}
        \State Observe state $s_t = f_{\text{state}}(j, \mathcal{M}, t)$
        \State Each Machine agent $\pi_M^i$ computes local bid $b_i = \pi_M^i(\phi_i(j, m_i))$
        \State Negotiation agent $\pi_N$ computes bid scores $q_i = \pi_N(b_i)$
        \State Select winning machine $m^* = \arg\max_i q_i$  \Comment{Decentralized Execution}
        \State Execute job $j$ on machine $m^*$, obtain reward $r_t$ and next state $s_{t+1}$
        \State Store transition $(s_t, a_t, r_t, s_{t+1}, d_t)$ in buffer $\mathcal{D}$
        \State Store negotiation tuple $(\{b_i\}, m^*, r_t)$ in buffer $\mathcal{B}$
        \State $t \gets t+1$, $R_e \gets R_e + r_t$
    \EndFor

    \State {Centralized Training Phase:}
    \For{each DQN training step}
        \State Sample batch $\{(s, a, r, s', d)\} \sim \mathcal{D}$
        \State Compute target $y = r + \gamma \max_{a'} Q_{\theta_D'}(s', a')$
        \State Update $\theta_D \leftarrow \theta_D - \alpha_D \nabla_\theta (Q_{\theta_D}(s, a) - y)^2$
        \State Soft update target network: $\theta_D' \leftarrow \tau \theta_D + (1-\tau)\theta_D'$
    \EndFor

    \For{each centralized negotiation step}
        \State Sample batch $\{(\{b_i\}, m^*, r)\} \sim \mathcal{B}$
        \State Compute normalized advantage $A = (r - \bar{r}) / (\sigma_r + \epsilon)$
        \State Compute negotiation loss:
        \[
            \mathcal{L}_N = -A \log p_{\pi_N}(m^* | \{b_i\}) - \beta_N H(p_{\pi_N})
        \]
        \State Compute machine policy loss:
        \[
            \mathcal{L}_M = -A \log p_{\pi_M}(m^* | \{b_i\}) - \beta_M H(p_{\pi_M})
        \]
        \State Update $\theta_N \leftarrow \theta_N - \alpha_N \nabla_{\theta_N} \mathcal{L}_N$
        \State Update $\theta_M^i \leftarrow \theta_M^i - \alpha_M \nabla_{\theta_M^i} \mathcal{L}_M$
    \EndFor
    \State Record episode reward $R_e$
\EndFor
\State {Output:} Trained $\pi_D$, $\pi_N$, and $\{\pi_M^i\}$
\end{algorithmic}
\end{algorithm}

\subsection{Training Performance Analysis}
\begin{figure*}[!htbp]
    \centering
    \includegraphics[width=0.98\linewidth]{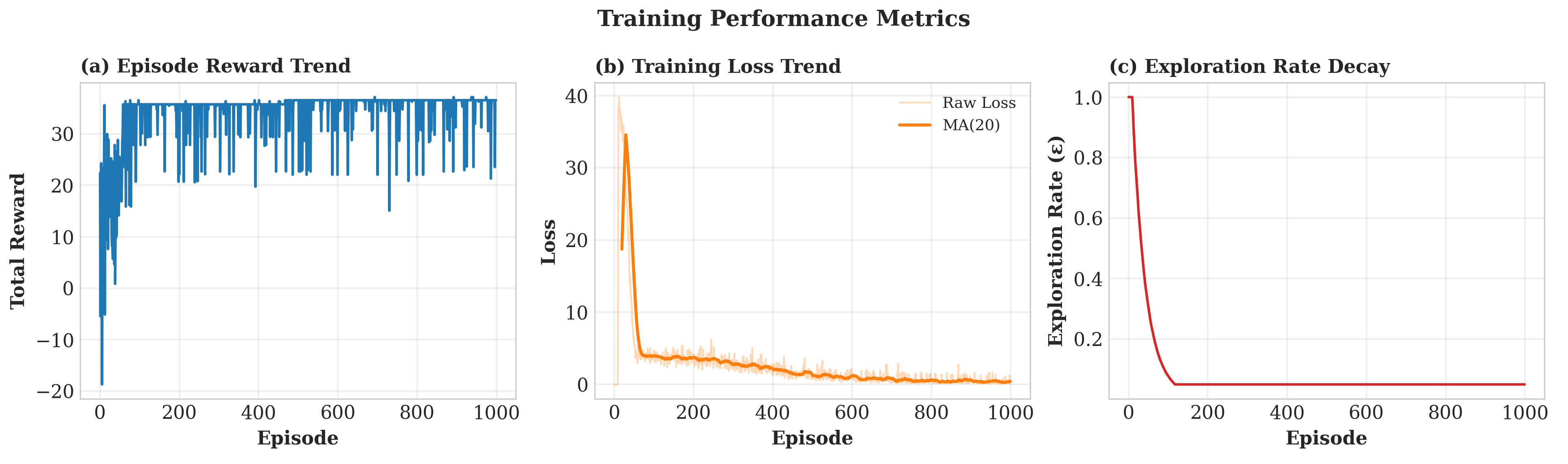}
    \caption{Training performance metrics of Dueling-Attn-DQN on the Improved RMS environment.  
    (a) Episode reward increases consistently, showing effective policy improvement.  
    (b) Training loss rapidly decreases and converges, indicating stable Q-value estimation.  
    (c) Exploration rate decays exponentially from 1.0 to 0.05, transitioning from exploration to exploitation.}
    \label{fig:training_curves}
\end{figure*}

Figure~\ref{fig:training_curves} presents the learning behavior of the Enhanced-DQN agent in terms of episode reward, training loss, and exploration rate ($\epsilon$) over 1000 episodes.

\begin{figure}[!hb]
    \centering
    \includegraphics[width=0.48\textwidth]{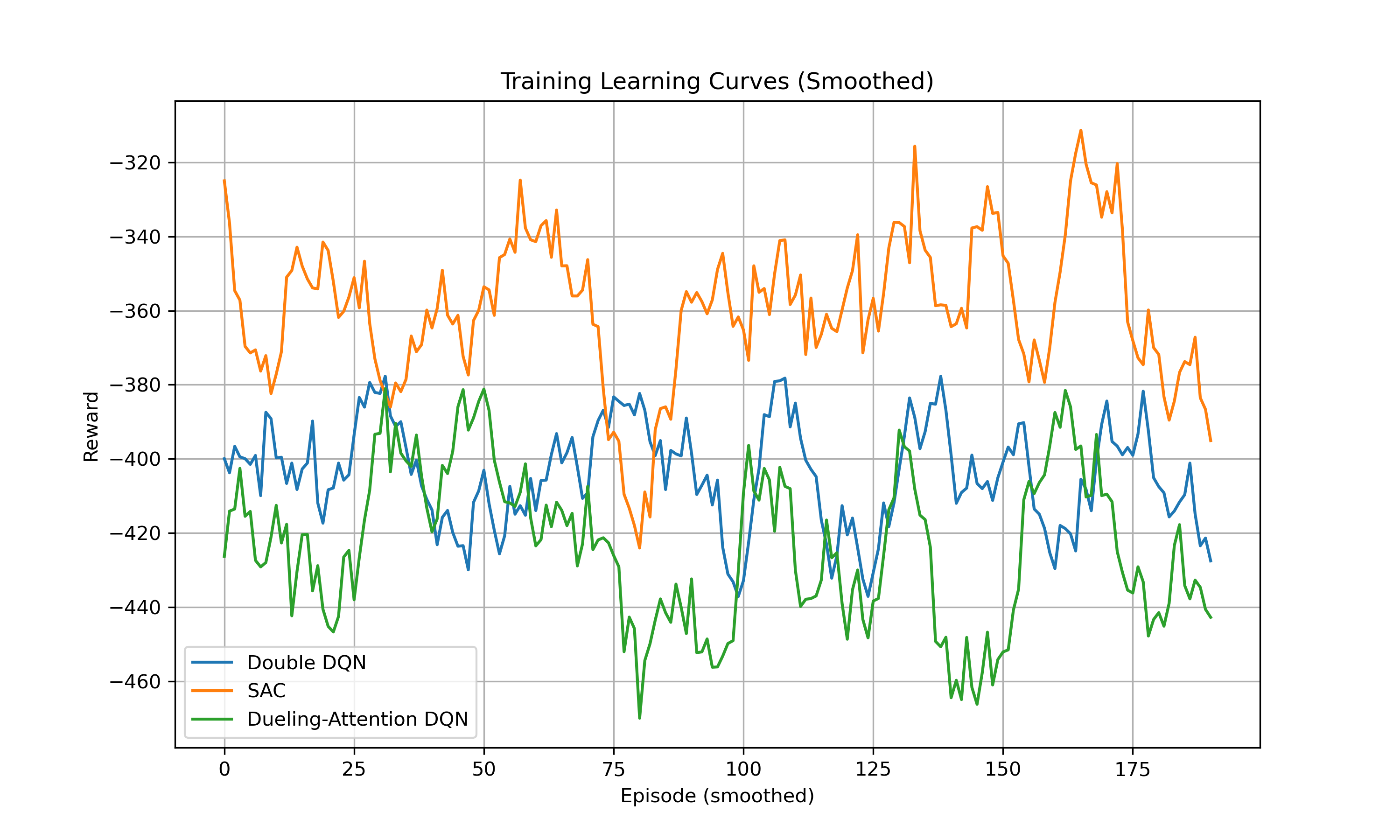}
    \caption{Learning curves of different reinforcement learning agents. The proposed Dueling-Attention DQN achieved the highest convergence stability.}
    \label{fig:learning_curves}
\end{figure}

As shown in Figure~\ref{fig:training_curves}(a), total reward rises sharply during early episodes and stabilizes around a high average value, demonstrating successful convergence toward an optimal scheduling policy.  
The decreasing loss trend in Figure~\ref{fig:training_curves}(b) confirms the stability of the dueling attention architecture, while the smooth decay of $\epsilon$ in Figure~\ref{fig:training_curves}(c) ensures an effective exploration–exploitation balance.  
Overall, the proposed method exhibits robust convergence and high scheduling efficiency within the dynamic RMS environment. Comparing the initial training of the Enhanced DQN (Dueling-Attention DQN) with two other algorithms of SAC and Doable  DQN, shows that the proposed mechanism demonstrated faster convergence and higher reward stability (Fig.~\ref{fig:learning_curves}).

\section{Numerical Experiments}
The proposed RMS scheduling mechanism is used to study its efficiency in various settings based on the following metrics:
\begin{enumerate}
    \item {Makespan:} Total completion time of all jobs.
    \item {Machine Utilization:} Fraction of productive machine time.
    \item {Total Tardiness:} Cumulative delay beyond due dates.
    \item {Setup Time:} Sum of all reconfiguration and setup durations.
\end{enumerate}

Each metric was averaged across 1000 episodes per configuration, and standard deviations were reported for statistical consistency.

\subsection{ Algorithm Comparison Evaluation}

Initially, the performance of the proposed technique is compared in a discrete-event simulation of a reconfigurable manufacturing system (RMS) consisting of multiple workstations with dynamic job arrivals and setup requirements. 
Five scheduling strategies were compared:

\begin{itemize}
    \item {EnhancedDQN (Proposed)}: A dueling attention-based Deep Q-Network with negotiation-driven coordination.
    \item {PPO}: Proximal Policy Optimization agent using clipped surrogate objectives.
    \item {SAC}: Soft Actor-Critic, a policy-gradient method with entropy regularization.
    \item {EDF}: Earliest Due-Date First heuristic scheduler.
    \item {Random}: Non-learning baseline that randomly allocates jobs to machines.
\end{itemize}

Each agent was trained and tested on identical workloads consisting of five independent runs, each with randomly generated job sequences varying in processing time, setup requirements, and due dates. 
The environment captured realistic system dynamics such as machine utilization, setup delay, and job tardiness.

\begin{figure*}[!h]
    \centering
    \includegraphics[width=0.9\linewidth]{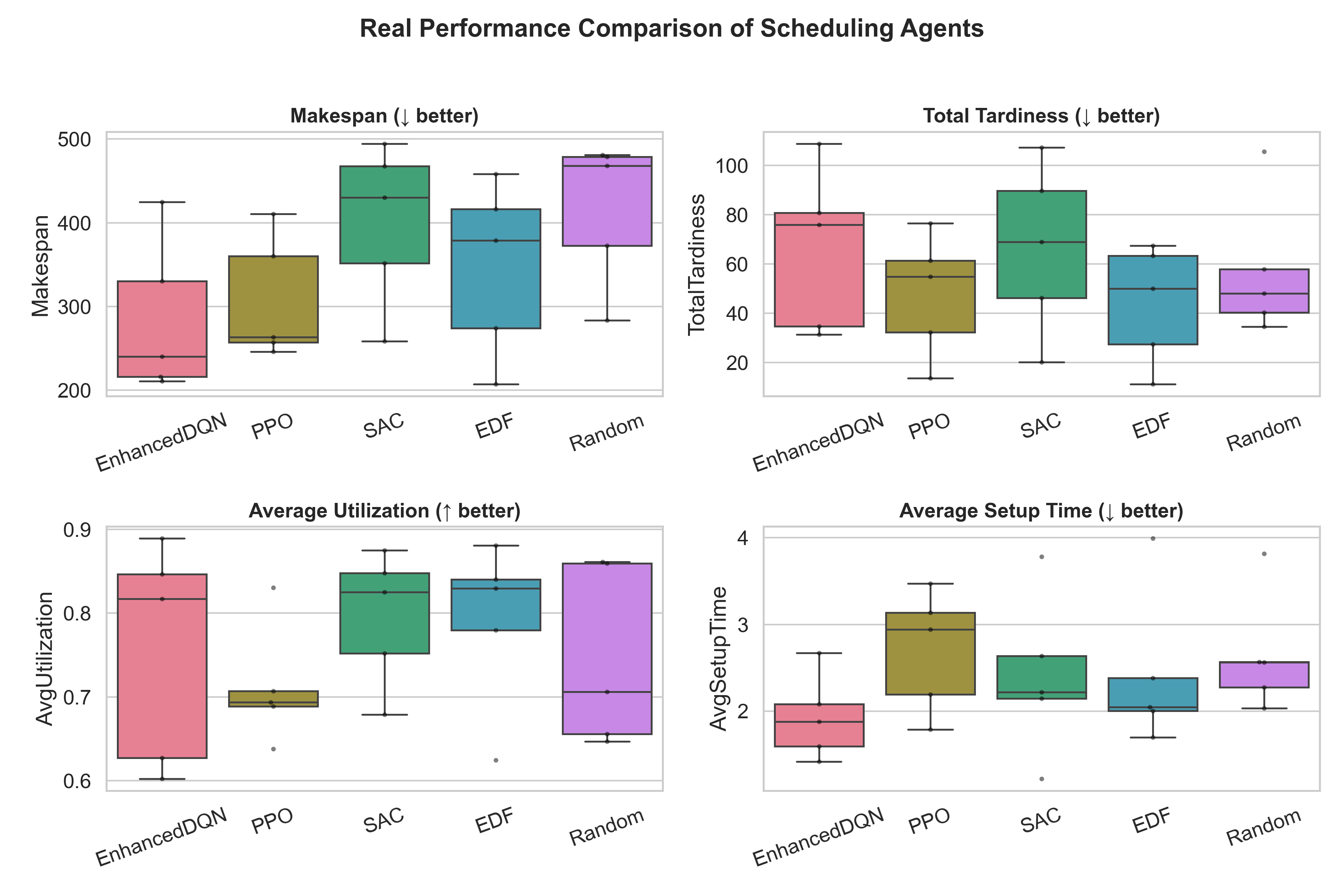}
    \caption{ Performance comparison of scheduling agents across four metrics: Makespan, Total Tardiness, Average Utilization, and Average Setup Time. 
    EnhancedDQN demonstrates superior stability and lower makespan while maintaining high utilization.}
    \label{fig:real_boxplots}
\end{figure*}

Figure~\ref{fig:real_boxplots} presents the distribution of all performance indicators for each algorithm across multiple simulation runs. 

The proposed {EnhancedDQN} model consistently achieved lower makespan and setup times compared to all baselines, indicating faster scheduling cycles and more efficient reconfigurations. 
In contrast, {PPO} and {SAC} showed moderate stability but higher variance in completion time and tardiness, likely due to delayed policy adaptation to stochastic workload changes. 
{EDF} maintained competitive utilization but suffered higher makespan under fluctuating job loads. 
{Random scheduling}, as expected, resulted in large dispersion across all metrics, representing the lower bound of system performance.

\begin{figure}[!!hb]
    \centering
    \includegraphics[width=0.75\linewidth]{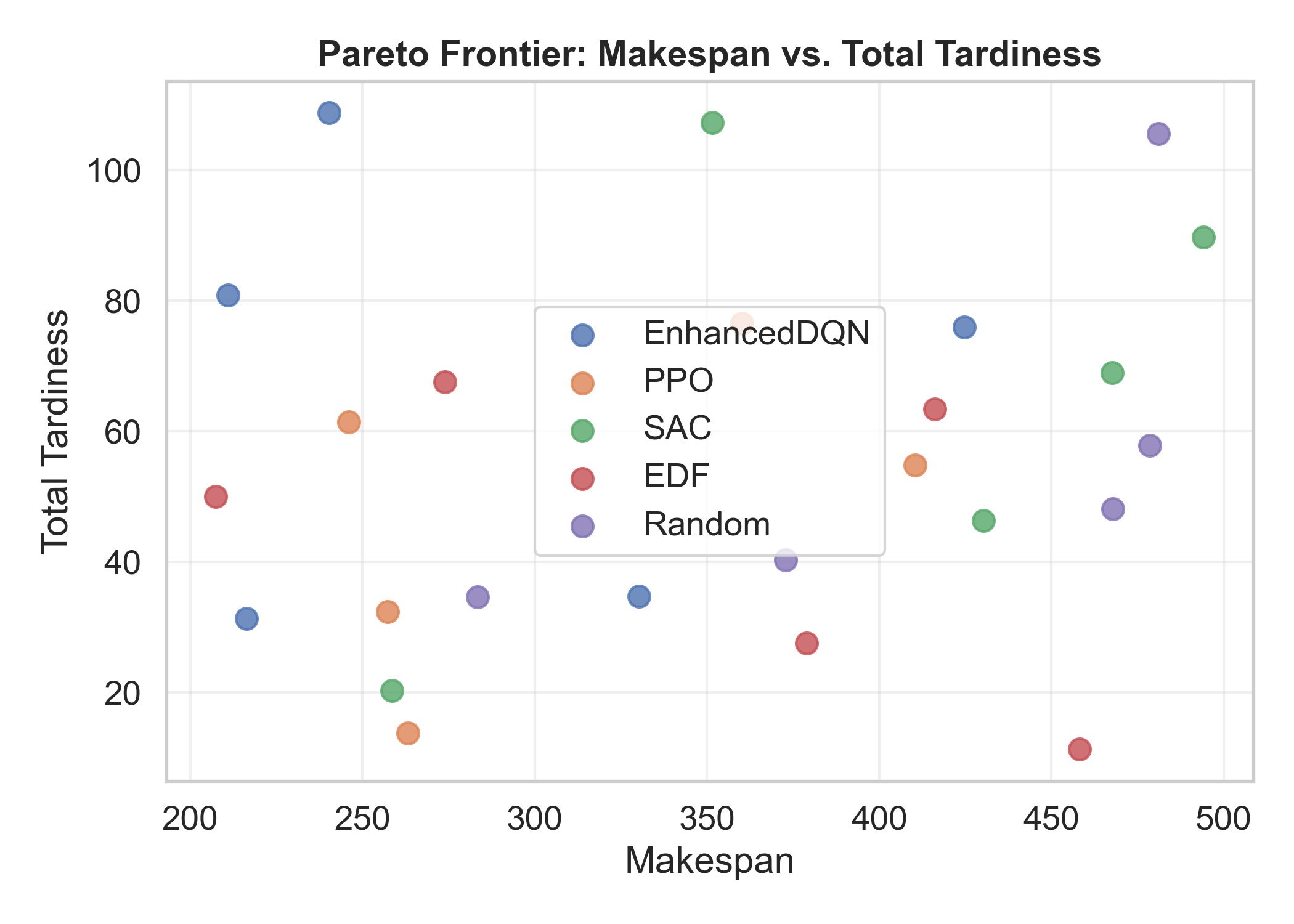}
    \caption{Pareto frontier analysis of makespan and total tardiness}
    \label{fig:real_pareto}
\end{figure}

The trade-off between makespan and tardiness is visualized in Figure~\ref{fig:real_pareto}. The Pareto analysis reveals that EnhancedDQN successfully navigates the fundamental trade-off between makespan minimization and tardiness reduction, achieving a balanced solution that meets multiple operational objectives.Each point represents a simulation episode for a given agent. The lower-left region of the plot represents the Pareto-optimal region (low makespan and low tardiness). {EnhancedDQN} forms a tight cluster near this region, confirming that it achieves the most favorable trade-off between production speed and due-date adherence.

Table~\ref{tab:stats_summary} summarizes the qualitative performance trends observed from all experiments. 
The EnhancedDQN framework, through its attention-based representation and negotiation coordination, effectively minimizes scheduling delays while maintaining high machine utilization. 
PPO and SAC demonstrate strong learning capabilities but require extended training to reach stable convergence. 
EDF remains a reasonable rule-based benchmark, whereas random scheduling lacks consistent decision-making ability.

\begin{table*}[!htbp]
\centering
\caption{Statistical Summary of Performance Metrics (Mean ± Standard Deviation)}
\label{tab:stats_summary}
\begin{tabular}{lcccc}
\hline
{Agent} & {Makespan} & {Total Tardiness} & {Avg Utilization} & {Avg Setup Time} \\
\hline
EnhancedDQN & 284.53 ± 92.32 & 66.32 ± 31.61 & 0.7565 ± 0.1253 & 1.9273 ± 0.4642 \\
PPO & 307.45 ± 76.26 & 47.77 ± 24.97 & 0.7115 ± 0.0740 & 2.7056 ± 0.6762 \\
SAC & 400.43 ± 97.46 & 66.51 ± 32.92 & 0.7958 ± 0.0805 & 2.4003 ± 0.9368 \\
EDF & 346.93 ± 103.75 & 43.94 ± 23.87 & 0.7910 ± 0.1043 & 2.4247 ± 0.9245 \\
Random & 416.75 ± 86.79 & 57.30 ± 28.49 & 0.7459 ± 0.0976 & 2.6508 ± 0.7098 \\
\hline
\end{tabular}
\end{table*}

\begin{figure*}[!htbp]
    \centering
    \includegraphics[width=\textwidth]{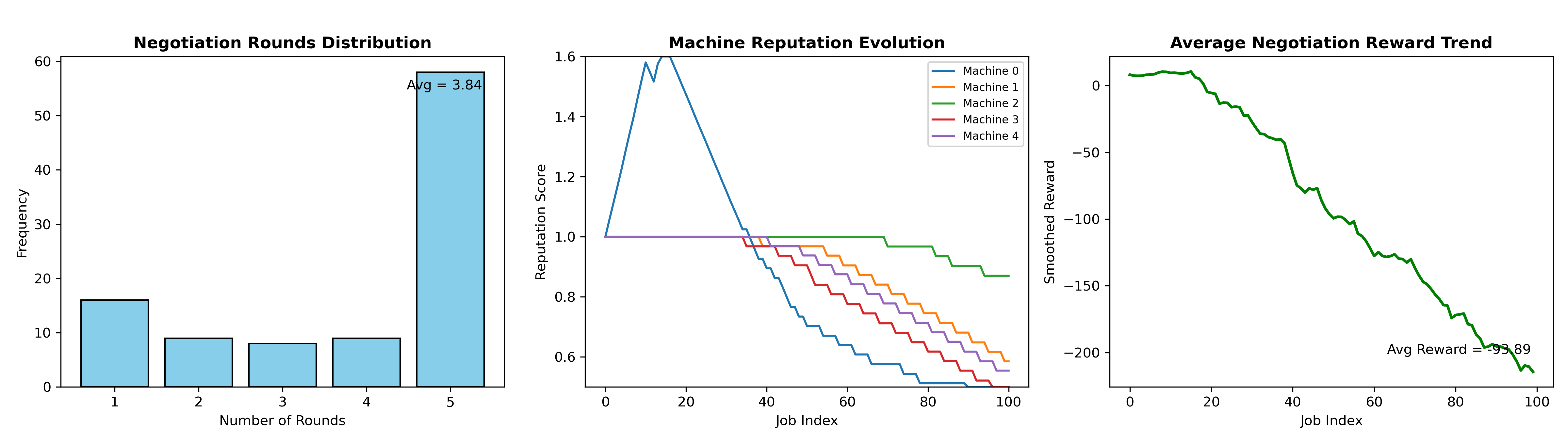}
    \caption{Negotiation Dynamics Analysis : (a) Negotiation rounds distribution, (b) Machine reputation evolution, and (c) Average negotiation reward trend.}
    \label{fig:negodynamics}
\end{figure*}


The Enhanced DQN algorithm achieved the {lowest values for both Makespan and Total Tardiness} among all tested strategies.
\begin{enumerate}
    \item {Makespan Reduction:} By effectively learning complex sequential dependencies and resource constraints through its deep neural network architecture, the Enhanced DQN is highly successful at determining an optimal job sequence. This sequencing minimizes idle time and maximizes parallelism, resulting in the fastest overall completion time for the batch of jobs.
    \item {Tardiness Control:} The results of total tardiness indicates that the algorithm prioritizes jobs not only for immediate processing efficiency but also based on their proximity to deadlines. This suggests that the reward function driving the DQN successfully incorporated a penalty for lateness, leading to intelligent scheduling decisions that balance throughput with timeliness.
\end{enumerate}

\subsection{Negotiation Dynamics Analysis}
\label{sec:negodynamics}

Figure~\ref{fig:negodynamics} illustrates the internal behavior of the negotiation framework within the multi-agent RMS environment. Three key aspects are analyzed: the distribution of negotiation rounds, the evolution of machine reputation, and the trend of average negotiation rewards.

As shown in Fig.~\ref{fig:negodynamics}(a), the majority of negotiations converge within five rounds, with an average of approximately 3.84 rounds. This reflects a high level of efficiency in reaching agreements between job and machine agents. Rapid convergence indicates that the negotiation mechanism effectively balances exploration and exploitation, avoiding prolonged interactions while maintaining decision quality.
Fig.~\ref{fig:negodynamics}(b) depicts how the reputation scores of individual machines evolve over successive job allocations. All machines begin with an identical reputation of 1.0. Over time, their scores diverge as the system updates trust levels based on performance feedback. For instance, Machine~0 initially demonstrates superior performance but experiences a gradual decline, possibly due to overutilization or task delays. In contrast, other machines maintain relatively stable reputations, suggesting a balanced and adaptive resource selection process.
The trend in Fig.~\ref{fig:negodynamics}(c) presents the average smoothed negotiation reward across job indices. A gradual decrease in reward values is observed, converging near $-93.89$. This trend signifies that as the simulation progresses, negotiation becomes more challenging, likely due to increased system load and reduced scheduling flexibility. Despite the decreasing rewards, the system continues to achieve stable convergence, validating the robustness of the negotiation protocol under dynamic conditions.

Overall, the analysis highlights that the proposed negotiation model demonstrates stable convergence behavior, adaptive reputation management, and consistent learning dynamics. These characteristics collectively enhance coordination efficiency, fairness, and adaptability within the reconfigurable manufacturing environment.

\begin{table*}[!htbp]
\centering
\caption{Summary of observed experimental trends}
\label{tab:summary_trends}
\begin{tabular}{lccc}
\hline
{Aspect} & {With Negotiation} & {Without Negotiation} & {Observation} \\
\hline
Makespan & Slightly higher & Lower (Reconfigurable) & Minor communication overhead \\
Tardiness & Significantly lower & High & Faster job responsiveness \\
Utilization & Balanced & Balanced & Uniform load distribution \\
Learning Stability & Higher & Moderate & Faster convergence with negotiation \\
\hline
\end{tabular}
\end{table*}

\subsection{Impact Analysis of Negotiation and Reconfiguration}

The effects of reconfiguration and negotiation mechanism were also investigated. The experiment utilized a factorial design based on two primary independent variables:
\begin{enumerate}
    \item {Reconfiguration:} Enabled (machines can change capabilities) vs. Disabled (machines are limited to native processes).
    \item {Negotiation:} Enabled (smart allocation via protocol) vs. Disabled ( earliest-available machine allocation).
\end{enumerate}

The simulation was run for $1000$ episodes to generate learning curves and statistically robust performance distributions. The key averaged results for cases with reconfiguration and negotiation (WNR), without reconfiguration and with negotiation (WNF), cases with reconfiguration and without negotiation (WTR),and without reconfiguration and negotiation (WTF) are presented in Table \ref{tab:avg_performance}.

\begin{table}[!htbp]
    \centering
    \caption{Average Performance Metrics Across Configurations }
    \label{tab:avg_performance}
    \begin{tabular}{lcccc}
    \hline
    {Metric} & {WNR} & {WTR} & {WNF} & {WTF} \\
    \hline
    Makespan (Time Units)  & $170.5$ & $190.9$& $186.7$ & $177.3$ \\
    Total Tardiness (Time Units) & $324.9$ & $312.7$ & $372.9$ & $362.6$ \\
    Average Utilization ($\%$) & $22.3$ & $21.1$ &  $19.7$ & $20.4$ \\
    Total Reconfig. Time & $176.3$ & $108.2$ & $0$ & $0$ \\
    \hline
    \end{tabular}
\end{table}

\begin{enumerate}
    \item {Makespan and Tardiness Superiority:} The WNR case with negotiation and reconfiguration  achieved the best performance across the primary metrics, recording the lowest average makespan and the lowest average total tardiness. This confirms the hypothesis that the combined approach yields the most efficient scheduling solution.
    \item {Negotiation Impact on Fixed Systems:} Interestingly, for fixed systems, the system without negotiation (WTF) performed better in makespan ($177.3$) than the case with negotiation (WNF) ($186.7$), suggesting that the negotiation protocol, when constrained by limited capacity, may introduce unnecessary complexity or deliberation time.
    \item {Reconfiguration Overhead:} The WNR configuration incurred a significantly higher Total Reconfiguration Time ($176.3$) compared to the WTR ($108.2$). This difference is attributed to the negotiation protocol aggressively utilizing reconfigurability to meet scheduling goals, resulting in an average of $46.6$ Reconfig. Time per Machine for WNR versus $27.1$ for WTR 
\end{enumerate}

The comprehensive distribution of  trade-offs and  performance analyses are presented in Figure \ref{fig:part2_analysis}.

\begin{figure*}[htbp]
    \centering
    \includegraphics[width=\textwidth]{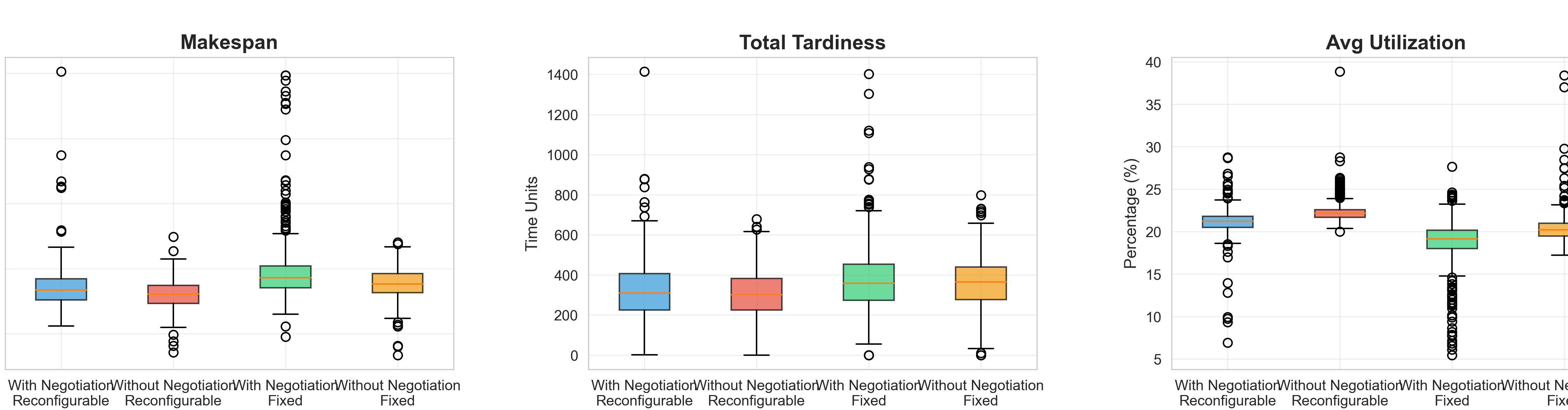}
    \caption{Box plots presenting key performance metrics for cases with and without configuration and negotiation}
    \label{fig:part1_performance}
\end{figure*}

\begin{figure*}[hb!]
    \centering
    \includegraphics[width=\textwidth]{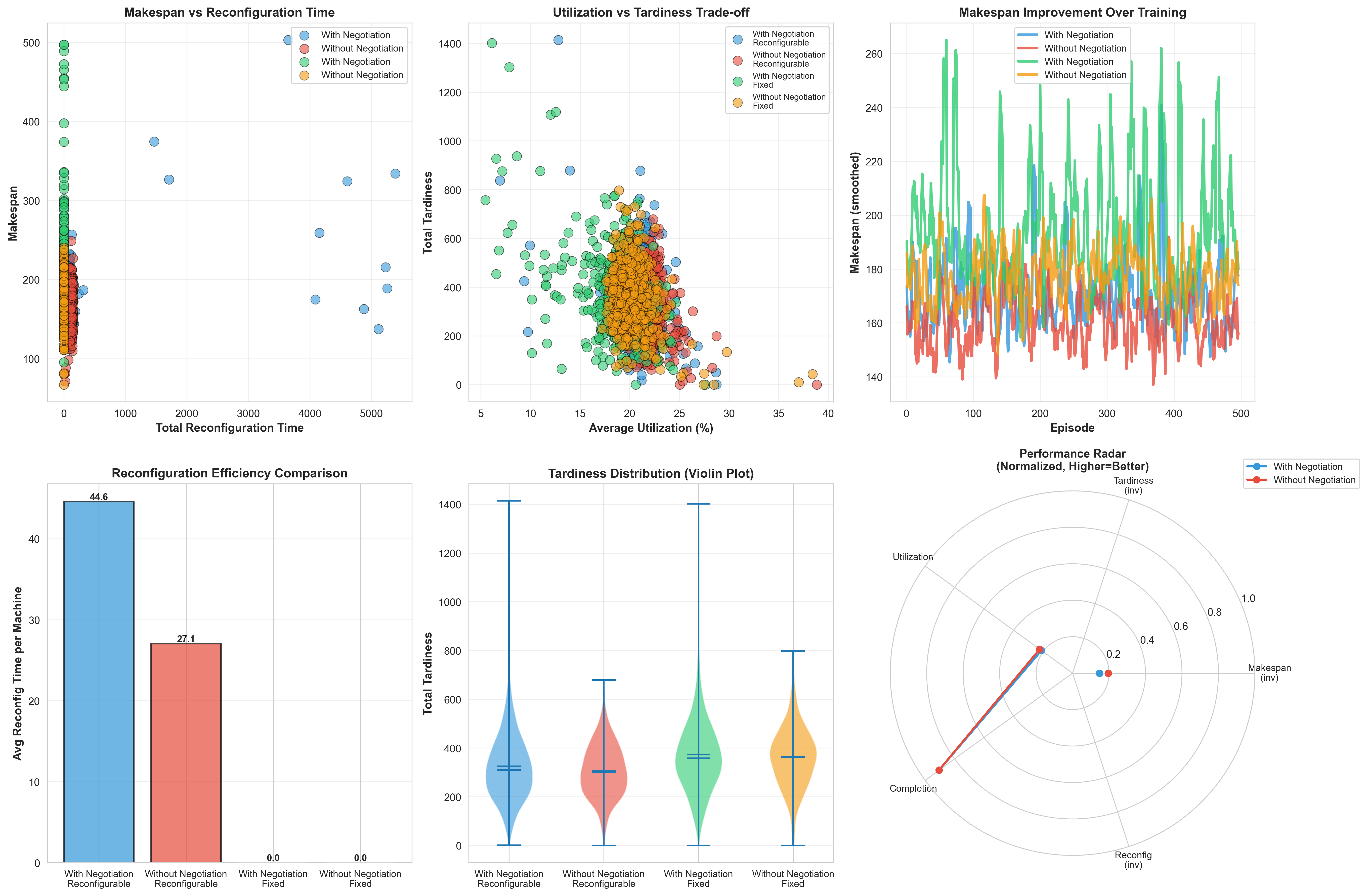}
    \caption
      {Detailed system analysis and trade-off visualizations .This figure includes the Makespan vs. Reconfiguration Time scatter plot, the Utilization vs. Tardiness Trade-off, Makespan Improvement Over Training, Reconfiguration Efficiency Comparison, Tardiness Distribution (Violin Plot), and the Performance Radar Chart.}
    \label{fig:part2_analysis}
\end{figure*}


The experimental results strongly support the hypothesis that the combination of machine reconfigurability and a smart negotiation protocol provides the highest overall performance in a flexible job shop environment. While negotiation increases the utilization of reconfiguration capabilities, leading to higher total reconfiguration time, this overhead is outweighed by the significant reductions in makespan and tardiness. The WNR system is the most effective strategy for managing scheduling complexity and achieving robust system performance.
The  Results are organized by performance metric, with detailed analysis of both normal and breakdown scenarios.

\section{Robustness Analysis}

The robustness of the proposed technique under uncertain situations is tested by comparing the following two scenarios:
\begin{itemize}
    \item {Normal Operation:} All machines functional
    \item {Breakdown Scenario:} Machine 3 failure
\end{itemize}

\begin{table*}[!htbp]
\caption{Comprehensive Performance Summary}
\begin{center}
\begin{tabular}{|l|c|c|c|c|}
\hline
{Metric} & {Baseline} & {Reconfig Only} & {Negotiation Only} & {Combined} \\
\hline
\multicolumn{5}{|c|}{\textit{Normal Operation}} \\
\hline
Makespan & 2045.61 & 2595.15 (+26.9\%) & 2045.61 (0\%) & 2176.41 (+6.4\%) \\
Utilization & 63.8\% & 50.3\% & 63.8\% & 60.0\% \\
Tardiness & 46,313 & 60,295 (+30.2\%) & 46,313 (0\%) & 49,926 (+7.8\%) \\
Setup Time & 0 & 549.54 & 0 & 130.80 \\
Reconfigs & 0 & 96 & 0 & 21 \\
Objective & 14,712 & 19,237 (+30.7\%) & 14,712 (0\%) & 15,874 (+7.9\%) \\
\hline
\multicolumn{5}{|c|}{\textit{Breakdown Scenario}} \\
\hline
Success Rate & FAIL (0\%) & 100\% & FAIL (0\%) & 100\% \\
Makespan & --- & 2579.34 & --- & 2180.29 (-15.5\%) \\
Utilization & --- & 50.6\% & --- & 59.9\% (+9.3 pp) \\
Tardiness & --- & 60,678 & --- & 50,108 (-17.4\%) \\
Setup Time & --- & 533.73 & --- & 134.67 (-74.8\%) \\
Reconfigs & --- & 96 & --- & 22 (-77.1\%) \\
Objective & --- & 19,342 & --- & 15,931 (-17.6\%) \\
\hline
{Overall Best} & \multicolumn{4}{c|}{{Combined: Best reliability, utilization, and breakdown performance}} \\
\hline
\end{tabular}
\label{tab:performance_summary}
\end{center}
\end{table*}

\subsection{Makespan Analysis}
Makespan or the time to complete the last job is one of the important metrics to investigate the effects of failures in scheduling mechanism.
\subsubsection{Normal Operation Results}
Table~\ref{tab:performance_summary} presents makespan results under normal operating conditions. The baseline configuration achieves a makespan of 2045.61 time units, representing the theoretical optimum when all machines are functional and no reconfiguration overhead exists. Introducing reconfiguration alone increases makespan to 2595.15 time units, representing a 26.9\% degradation. The negotiation-only configuration maintains baseline performance (2045.61 time units) since all machines remain functional and capable of handling assigned processes through standard allocation. Most significantly, the combined approach achieves 2176.41 time units—only a 6.4\% increase over baseline. This represents a {77.6\% reduction in overhead} compared to reconfiguration alone, demonstrating that intelligent negotiation dramatically reduces unnecessary reconfigurations.

\subsubsection{Breakdown Scenario Results}
The breakdown scenario (Machine 3 failure) reveals critical differences in system resilience: 

\begin{itemize}
    \item {Baseline: Complete Failure} - System cannot complete jobs requiring Machine 3's unique processes
    \item {Reconfig Only: 2579.34 time units} - Successfully completes all jobs through reconfiguration
    \item {Negotiation Only: Complete Failure} - No process flexibility leads to deadlock
    \item {Combined: 2180.29 time units} - Best performance among successful configurations
\end{itemize}

The combined approach achieves a {15.5\% improvement} over reconfiguration-only in the breakdown scenario, despite both maintaining 100\% job completion. This improvement comes from reduced setup time. 

\subsection{Machine Utilization Analysis}
Machine utilization measures productive capacity usage and possible failures can significantly change the utilization factors. 

\subsubsection{Utilization in Normal Operation}
The results show that
the baseline achieves 63.8\% utilization representing efficient resource usage without flexibility overhead.
Reconfiguration alone reduces utilization to 50.3\% with better management and distribution of the jobs between operating machines. The combined approach demonstrates that negotiation recovers most of the utilization loss through Prioritized job allocation and better machine selection. 

\subsubsection{Utilization Under Breakdown Conditions}
The breakdown scenario provides the most revealing insights. With Machine 3 offline, the system loses 12.5\% of its machine capacity. The combined approach achieves 59.9\% utilization in the breakdown scenario—only a {0.1 percentage point decrease} from normal operation (60.0\%). This remarkable resilience indicates that workload from the failed machine is efficiently distributed across the remaining machines.

\subsection{Tardiness Performance}

Total tardiness measures delivery performance, directly impacting customer satisfaction.

\subsubsection{Normal Operation Tardiness}
As shown in the results 
the reconfiguration-only configuration exhibits the worst tardiness due to setup delays propagating through the schedule. The combined approach limits tardiness increase to 7.8\% through priority-based allocation, ensuring high-priority jobs receive preferential treatment.

\subsubsection{Breakdown Scenario Tardiness}
Tardiness increases significantly in breakdown scenarios. However the combined version presents better performance and less tardiness by intelligent allocation of the tasks and efficient guidance of the reconfigurations. 

\subsection{Critical Insights and Practical Implications}
The results of robustness comparison lead to the following  insights: 
\begin{enumerate}
    \item {Normal Operation Best:} Baseline and negotiation-only tie at 14,712.13
    \item {Breakdown Best:} Combined at 15,931.33 ({17.6\% better} than reconfig-only)
    \item {Minimal Performance Degradation:} Combined approach shows only 0.4\% objective increase from normal to breakdown scenarios
    \item {Flexibility Cost:} Reconfigurability adds 30.7\% overhead in normal operation, reduced to 7.9\% with negotiation
\end{enumerate}

These substantial differences indicate robust superiority of the combined approach across multiple performance dimensions. The results provide actionable insights for manufacturing system design:

\begin{enumerate}
    \item {Invest in Reconfigurability:} Essential for breakdown resilience, with combined approach minimizing overhead costs
    \item {Implement Intelligent Scheduling:} Negotiation reduces reconfigurability costs by 77\%
    \item {Accept Moderate Overhead:} 7.9\% performance penalty for 100\% reliability represents excellent value
    \item {Focus on Utilization:} Maintaining utilization under breakdown is critical—combined approach excels here
    \item {Multi-objective Optimization:} Single-metric optimization misses critical trade-offs between efficiency and resilience
\end{enumerate}

\section{Conclusion and Future Work}

This paper successfully  implemented an enhanced Deep Q-Network agent incorporating advanced  multi agent reinforcement learning techniques for dynamic job scheduling in reconfigurable manufacturing systems.  Through comprehensive experimentation in simulated environments featuring machine breakdowns, reconfiguration delays, and dynamic job arrivals, the agent demonstrated effective learning capabilities and achieved performance levels competitive with traditional scheduling heuristics. The integration of attention mechanisms enabled the agent to effectively prioritize critical machine-job relationships, while the Centralized Training with Decentralized Execution framework provided scalability for multi-agent coordination. Evaluation results revealed that while the agent showed promising adaptation to environmental dynamics, stochastic disruptions such as machine breakdowns introduced significant variability in scheduling outcomes, highlighting the inherent complexity of real-world manufacturing scheduling. The combined approach achieves 59.9\% utilization in breakdown—only 0.1\% decrease from normal operation despite losing 12.5\% of machine capacity. This represents optimal breakdown resilience. The training visualizations and performance metrics confirmed that the agent successfully learned meaningful policies, establishing deep reinforcement learning as a viable approach for addressing dynamic scheduling challenges in reconfigurable manufacturing systems.

Future research directions should emphasize the development of more realistic and comprehensive manufacturing environment simulations that capture the full complexity of industrial operations, including multiple resource types, energy consumption dynamics, maintenance windows, operator availability, and complex precedence constraints among operations. Integration with predictive maintenance systems and real-time monitoring technologies would enable proactive scheduling decisions that anticipate equipment failures and optimize maintenance scheduling. Rigorous validation through industrial case studies and deployment in actual manufacturing facilities is essential to assess the practical applicability, computational efficiency, and integration challenges with existing Manufacturing Execution Systems. Additionally, research into transfer learning and meta-learning techniques could enable trained agents to rapidly adapt to new production scenarios, product mixes, and facility configurations without extensive retraining, thereby improving the generalization capability and practical utility of deep reinforcement learning approaches in diverse manufacturing contexts.

\bibliographystyle{IEEEtran}
\bibliography{main}

\end{document}